\documentclass[11pt,a4paper]{article}

\usepackage{latexsym}
\usepackage{amssymb}
\usepackage{amsmath}
\usepackage[authoryear]{natbib}

\newif\ifpdf
\ifx\pdfoutput\undefined
   \pdffalse
\else
   \pdfoutput=1
   \pdftrue
\fi

\ifpdf
    \usepackage{color}
    \definecolor{myred}{rgb}{0.5,0,0}
    \definecolor{myblue}{rgb}{0,0,0.75}
    \definecolor{mygreen}{rgb}{0,0.5,0}
   \usepackage[pdftex,
               pdftitle={Calculating Value-at-Risk contributions in CreditRisk+},
               pdfsubject={CreditRisk+ model},
               pdfkeywords={CreditRisk+; Value-at-Risk (VaR); risk contribution; conditional expectation},
               pdfauthor={Hermann Haaf; Dirk Tasche},
               pdfstartview=FitBH,
               breaklinks=true,
               colorlinks=true,
               citecolor=myred,
               linkcolor=myblue,
               urlcolor=mygreen]{hyperref}
\else
   \usepackage{hyperref}
\fi

\newtheorem{theorem}{Theorem}[section]

\newtheorem{remark}[theorem]{Remark}

\newtheorem{corollary}[theorem]{Corollary}
\newtheorem{assumption}[theorem]{Assumption}

\numberwithin{equation}{section}

\begin{document}
\title{Calculating Value-at-Risk contributions in CreditRisk$^+$%
\thanks{The authors thank Claudio Baraldi, Isa Cakir, Alexandre Kurth, Frank Lehrbass, Joachim Renz and Armin Wagner
for helpful suggestions and discussions about this paper.}
}
\author{
Hermann Haaf\thanks{Commerzbank AG, Risk Control (ZRC), Methods \&
Policies, 60261 Frankfurt/Main, Germany; E-mail:
\hbox{Hermann.Haaf@Commerzbank.com}} \hspace{1cm} Dirk
Tasche\thanks{RiskLab Switzerland, D-MATH, ETH Zentrum, 8092
Z\"urich, Switzerland;
E-mail: tasche@math.ethz.ch}%
}

\date{First version: November 19, 2001\\
This update: February 28, 2002} \maketitle

\begin{abstract}
Credit Suisse First Boston (CSFB) launched in 1997 the model
\emph{CreditRisk$^+$}
which aims at calculating the loss distribution of a credit portfolio on the basis of a methodology
from actuarial mathematics.
Knowing the loss distribution, it is possible to determine
quantile-based values-at-risk (VaRs) for the portfolio. An open question
is how to attribute \emph{fair} VaR contributions to the credits or loans
forming the portfolio. One approach is to define the contributions
as certain conditional expectations. We develop an algorithm for the calculations
involved in this approach. This algorithm can be adapted for computing
the contributions to the portfolio Expected Shortfall (ES).
\end{abstract}

\section{Introduction}
The
CreditRisk$^+$ model \citep{CRplus}
has found wide-spread applications for the measurement of risk in credit portfolios.
Reasons for the success of the model might be its
free availability, the speed of calculations based on it due to fact that no
Monte Carlo simulations need to be performed, and its detailed documentation.
The basic idea in the model is to apply a methodology from actuarial
mathematics.

As soon as the loss distribution has been computed, it is an easy task to calculate the
portfolio value-at-risk (VaR). A further step might be to perform a risk diagnostics
in the spirit of \citet{L96}. Such a diagnostics presupposes that the total portfolio
risk can be attributed to the portfolio components in a way that detects the impact on
the portfolio risk by the components.

A first suggestion for determining such VaR contributions was made in the
CreditRisk$^+$
documentation
\citetext{based on a decomposition of the portfolio variance \citep[see][]{CRplus}; \citealp[cf.~also][]{BKWW99}}.
Further proposals can be found
in
\citet[][sec.~5]{L01}
or
\citet{MBT01}.
In both these latter proposals the VaR contributions are motivated as approximations to partial derivatives of VaR.
Here we follow an alternative approach which relies on the fact that under certain continuity
assumptions the partial derivatives of VaR with respect to the weights of the assets are
just conditional expectations. This leads to an alternative definition of VaR contributions
(see (\ref{eq:7})) which might be useful for checking the results according to the approaches
in \citet{L01} or \citet{MBT01}. Moreover, this definition can easily be adapted for the case
of Expected Shortfall (ES) (see (\ref{eq:24})) as risk measure.

This text is based on the representations of the CreditRisk$^+$ model
as given
by \citet{L01} and \citet{G01}.
In section \ref{sec:abstr-descr-probl}, we introduce some basic features of the
CreditRisk$^+$ model and provide motivation for the kind of VaR contributions
to be discussed below. Then, in section \ref{sec:an-algorithm-var}, we study the model in a
more detailed  manner in order to derive the main result of this paper, Corollary \ref{co:2}.
It states
that once it is possible to calculate the probability masses of the
loss distributions, the computation of VaR contributions can be performed
essentially the same way. This result is adapted to the case of ES in section
\ref{sec:contr-expect-shortf}.
We conclude with a short summary of the results.

\section{Abstract description of the problem}
\label{sec:abstr-descr-probl}

\paragraph{The key idea in CreditRisk$^\mathbf{+}$.}
Consider a portfolio with $n$ loans or credits. Default or non-default
of loan $i$, $i = 1, \ldots, n$, by a fixed time horizon is indicated by the random variable
$D_i$ which can only take the values $1$ for default or $0$ for non-default.
Denote by $\nu_i$ the \emph{exposure} of loan $i$, measured as a positive integer multiple of a certain
currency unit. $p_i = \mathrm{P}[D_i =1] \in (0,1)$ is the \emph{default probability} of
loan $i$. The \emph{realized} total loss $L_0$ of the portfolio is then given as
\begin{equation}
  \label{eq:1}
L_0 \ = \ \sum_{i=1}^n \nu_i\,D_i.
\end{equation}
The following assumption is quite common for the purpose of dependency modeling in a portfolio
of possible dependent assets
\citep[see][]{FM01}. Suppose that all the random variables under consideration are defined on
the underlying probability space $(\Omega, \mathcal{F}, \mathrm{P})$.

%
\begin{assumption}
  \label{assump:key-idea-credit}
There is a $\sigma$-algebra $\mathcal{A} \subset \mathcal{F}$ such that the default indicators $D_1, \ldots, D_n$ are
independent conditional on $\mathcal{A}$.
\end{assumption}
We will see examples for the choice of $\mathcal{A}$ in section \ref{sec:an-algorithm-var}. $\mathcal{A}$
may be regarded as the collection of \emph{systematic} factors influencing the portfolio.

If the default probabilities $p_i$ are small, the total loss distribution might not change too much
when the Bernoulli variables $D_i$ in (\ref{eq:1}) are replaced by random variables $N_1, \ldots , N_n$ with
non-negative integer values that conditional on $\mathcal{A}$ are independent and Poisson-distributed
with $\mathcal{A}$-measurable intensities $R_1, \ldots, R_n > 0$ such that
\begin{equation}
  \label{eq:2}
\mathrm{E}[R_i]\ =\ p_i,\quad i = 1,\ldots,n.
\end{equation}
In the CreditRisk$^+$ model, one does not compute the distribution of the random variable $L_0$ from (\ref{eq:1})
but the distribution of the random variable
\begin{equation}
  \label{eq:3}
L \ = \ \sum_{i=1}^n \nu_i\,N_i
\end{equation}
which, of course, is only an approximation to the true distribution.

\paragraph{Which risk contributions should be considered?} For a fixed (close to 1) level $\delta \in (0,1)$,
\begin{equation}
  \label{eq:4}
\rho(L) \ = \ q_\delta(L)\ = \ \inf\{l \in \mathbb{R}:\, \mathrm{P}[L \le l] \ge \delta \}
\end{equation}
is a widespread measure of portfolio loss risk. It is called \emph{value-at-risk (VaR)} at level $\delta$ of $L$.
In order to identify
sources of particularly high risk in the portfolio, it is interesting to
decompose $\rho(L)$ into a sum $\sum_{i=1}^n \rho_i(L)$ where the
\emph{risk contributions} $\rho_i(L)$ should correspond in some sense to the
single losses $\nu_i\,N_i$.

\citet{L96}
suggested the decomposition
\begin{equation}
  \label{eq:5}
\rho(L) \ =\ \sum_{i=1}^n \frac{\partial \rho}{\partial h} \bigl( h\,\nu_i\,N_i + L\bigr)
\bigg|_{h=0}.
\end{equation}
Decomposition (\ref{eq:5}) holds whenever the risk measure $\rho$ is positively homogeneous
(i.e. $\rho(h\,L) = h\,\rho(L)$ for $h >0$) and differentiable. Unfortunately, the distribution
of the portfolio loss $L$, specified by (\ref{eq:3}), is purely discontinuous. Therefore, the derivatives
of $q_\delta(L)$ in the sense of (\ref{eq:5}) will either not exist or be $0$.

Nonetheless, for a large portfolio even the positive probabilities $\mathrm{P}[L =l]$ will be rather small.
On an appropriate scale, the distribution of $L$ will therefore appear
``almost'' continuous.
\citet{T99}
showed
\vspace{-2ex}
\begin{itemize}
\item that \citeauthor{L96}'s proposal (\ref{eq:5})
is the only decomposition which is compatible with
RORAC (Return on risk-adjusted capital) based portfolio management, and
\item that in case $\rho(L) = q_\delta(L)$ under certain continuity assumptions on the distribution of the
asset losses one would have
\begin{equation}
  \label{eq:6}
\frac{\partial q_\delta}{\partial h} \bigl( h\,\nu_i\,N_i + L\bigr)\bigg|_{h=0} \ =\
\nu_i\,\mathrm{E}[N_i\,|\, L = q_\delta(L)],\quad i = 1, \ldots, n.
\end{equation}
\end{itemize}
Note that under the assumptions of
\citet{T99}
the event $\{L = q_\delta(L)\}$ has probability $0$. Hence, the conditional
expectation in (\ref{eq:6}) has to be understood in the non-elementary sense
\citep[see][ch.~4]{D96}.

\citeauthor{L96}'s and \citeauthor{T99}'s considerations
suggest the definition of
\begin{equation}
  \label{eq:7}
\rho_i(L)\ = \ \nu_i\,\mathrm{E}[N_i\,|\, L = q_\delta(L)],\quad i = 1, \ldots, n,
\end{equation}
as VaR contributions of the loans $i = 1, \ldots, n$ in the portfolio.
Since $\mathrm{P}[L=q_\delta(L)]$ is positive by definition (\ref{eq:4}) of
VaR, the conditional expectation in (\ref{eq:7}) is elementary in the sense
that
\begin{equation}
  \label{eq:7a}
\mathrm{E}[N_i\,|\, L = q_\delta(L)]\ = \
\frac{\mathrm{E}[N_i\,\mathbf{1}_{\{L = q_\delta(L)\}}]}{\mathrm{P}[L=q_\delta(L)]},
\end{equation}
where, as usual, the indicator function $\mathbf{1}_A$ denotes a random
variable with value $1$ on the event $A$ and value $0$ on the complement of $A$.

\section{An algorithm for the VaR contributions}
\label{sec:an-algorithm-var}

The aim with this section is to derive expressions for the VaR contributions
defined by (\ref{eq:7}) which can be evaluated by the usual output of CreditRisk$^+$
or at least by minor modifications of this output.

In order to derive an expression for the expectation in (\ref{eq:7}) which can be numerically evaluated,
we have to specify the $\sigma$-algebra from Assumption \ref{assump:key-idea-credit}
and the stochastic intensities $R_1, \ldots, R_n$ from (\ref{eq:2}). This specification
constitutes what is commonly called the CreditRisk$^+$ model.
Recall that the Gamma-distribution with
parameters $(\alpha, \beta) \in (0, \infty)^2$ is defined by the density $f_{\alpha, \beta}$
with
\begin{equation}
  \label{eq:12}
f_{\alpha, \beta}(x) \ =\ \frac{x^{\alpha-1}}{\beta^\alpha\,\Gamma(\alpha)}\,e^{-x/\beta},\quad x >0.
\end{equation}
\vspace{-2ex}
\begin{assumption}
  \label{as:2}
The stochastic intensities $R_i$ of the $N_i$ in (\ref{eq:3}), $i = 1, \ldots, n$, are given as
\begin{equation}
  \label{eq:11}
R_i \ = \ \sum_{j=0}^k r_{j,i}\,S_j,
\end{equation}
where $r_{j,i} \ge 0$, $\sum_{j=1}^k r_{j,i} > 0$, $\sum_{i=1}^n r_{j,i} = 1$, $S_0 \ge 0$ is a constant,
and the $S_1, \ldots, S_k$ are independent and
Gamma-distributed with parameters $(\alpha_j, \beta_j)\in (0,\infty)^2$, $j=1, \ldots, k$. The $\sigma$-algebra
$\mathcal{A}$ is the $\sigma$-algebra generated by $(S_1, \ldots, S_k)$, i.e. $\mathcal{A} =
\sigma(S_1, \ldots, S_k)$. Eq.~(\ref{eq:2}) holds.
\end{assumption}
Usually, the variables $S_j$ are interpreted as systematic default intensities which are characteristic
for sectors, countries, or branches. Equation (\ref{eq:11}) then mirrors the fact that a firm may be influenced
by the evolutions in several sectors. Since $\mathrm{E}\bigl[S_j\bigr] = \alpha_j\,\beta_j$, $j = 1, \ldots, k$,
under Assumption \ref{as:2} eq.~(\ref{eq:2}) is equivalent to
\begin{equation}
  \label{eq:9}
p_i \ =\ r_{0,i}\,S_0 + \sum_{j=1}^k r_{j,i}\,\alpha_j\,\beta_j,\quad i = 1, \ldots, n.
\end{equation}
\begin{samepage}
\begin{remark}\
  \label{rm:1}
\nopagebreak[4]
\vspace{-2ex}
\begin{enumerate}
\item In practice, it is unlikely to have a credit portfolio parameterized like in Assumption \ref{as:2}.
In \citet[][sec.~A12]{CRplus} and \citet{L01}, for instance, the given data are the default probabilities
$p_i$, $i = 1, \ldots, n$, and default volatilities $\sigma_i$, $i = 1, \ldots, n$, for each credit as well
as factor weights $\theta_{j,i}$, $j = 0, \ldots, k$, $i= 1, \ldots, n$ which measure to which degree credit $i$
is exposed to factor $j$ and satisfy $\sum_{j=0}^k \theta_{j,i} =1$, $i = 1, \ldots, n$.
In this case, the $r_{j,i}$ are defined by
\begin{subequations}
  \begin{equation}
    \label{eq:10}
r_{j,i} \ =\ \frac{p_i\,\theta_{j,i}}{\mu_j},
  \end{equation}
with $\mu_j = \sum_{i=1}^n p_i\,\theta_{j,i} = \mathrm{E}\bigl[S_j\bigr]$. Hence,
$S_0$ is given by $S_0 = \mu_0$. It is quite common to specify also the variances $\tau^2_j$ of the factors $S_j$,
$j = 1, \ldots, k$ in order to determine the values of $(\alpha_j, \beta_j)$. Since $\mathrm{E}\bigl[S_j\bigr] =
\alpha_j\,\beta_j$ and $\mathrm{var}\bigl[S_j\bigr] = \alpha_j\,\beta_j^2$, we have, once the expectations $\mu_j$ and
the variances $\tau_j^2$ are known,
\begin{equation}
  \label{eq:17}
  \begin{split}
\alpha_j&\ = \ \frac{\mu_j^2}{\tau_j^2},\\
\beta_j & \ = \ \frac{\tau_j^2}{\mu_j}.
  \end{split}
\end{equation}
In \citet{CRplus}, the choice $\tau_j = \sum_{i=1}^n \theta_{j,i}\,\sigma_i$ is suggested in order to get a link
between the default volatilities $\sigma_i$ and the factor variances $\tau_j$. However, this approach tends to
underestimate the factor variances \citep[see][]{KL01}. Therefore \citeauthor{KL01} propose to calculate the
factor variances by
\begin{equation}
  \label{eq:20}
\tau_j\ =\ \sum_{i=1}^n \sqrt{\theta_{j,i}}\,\sigma_i, \quad j = 1, \ldots, k.
\end{equation}
\end{subequations}
%
%
\item In contrast to \citet{BKWW99} and \citet{G01}, we do \emph{not} assume $\mathrm{E}[S_j] =1$ in (\ref{eq:11}).
Indeed, this would imply $\alpha_j \beta_j = 1$ and, as a consequence, would render the formulation of the
main result in Corollary \ref{co:2} quite difficult. In a self-explanatory mixture of our notions
and the notions in \citet{G01}, \citeauthor{G01}'s equation (1) and (\ref{eq:11}) above are related by
\begin{equation}
  \label{eq:19}
  \begin{split}
    r_{j,i} & = \frac{w_{i j}\,\bar{p}_i}{\sum_{l=1}^n w_{l j}\,\bar{p}_l},\quad i = 1, \ldots,n, j = 0, \ldots, k,\\[1ex]
\mathrm{E}[S_j] & = \sum_{l=1}^n w_{l j}\,\bar{p}_l, \quad  j = 0, \ldots, k.
  \end{split}
\end{equation}
\end{enumerate}
\end{remark}
\end{samepage}
From Assumption~~\ref{as:2}, we obtain the following representation of
the generating function
$g(z) = \mathrm{E}[z^L]$ of the distribution of $L$ \citep[cf.][]{G01, L01}
\begin{subequations}
\begin{equation}
  \label{eq:13}
g(z) \ = \ \exp\left(S_0 \bigl({\textstyle\sum\limits_{i=1}^n} r_{0,i}\,z^{\nu_i}-1\bigr)\right)
\prod_{j=1}^k \left({1 +\beta_j - \beta_j\,{\textstyle\sum\limits_{i=1}^n} r_{j,i}\, z^{\nu_i}}\right)^{-\alpha_j}.
\end{equation}
Appealing to the binomial series it is clear that the radius of
convergence $\rho_g$ of the power series representation of $g$ is
given by
\begin{equation}
  \label{eq:31}
\rho_g \ =\ \sup\bigl\{ z > 0 :\, \sum_{i=1}^n r_{j,i}\, z^{\nu_i} < 1 + \frac 1 {\beta_j},\, j= 1, \ldots, k\bigr\}.
\end{equation}
Obviously, we have $\rho_g >1$, i.e. there are numbers $z >1$ for
which the power series of $g$ converges.
\end{subequations}
From (\ref{eq:13}), the exact distribution of the portfolio loss
$L$ can be successively determined \citep[see][A10]{CRplus} by
means of a recurrence relation.

More important here, we can derive from (\ref{eq:13}) the generating functions of the sequences
$l \mapsto \mathrm{E}\bigl[ N_i\,\mathbf{1}_{\{L = l\}}\bigr], i = 1, \ldots, n$.
\begin{theorem}
  \label{th:1}
Let $N_1, \ldots, N_n$ satisfy Assumption \ref{as:2}.
Define $L$ by (\ref{eq:7a}), $g$ by (\ref{eq:13}), and $\rho_g$ by (\ref{eq:31}). Then the functions
$f_i: (0,\rho_g) \to \mathbb{R}$, $i= 1, \ldots, n$ with
\begin{subequations}
\begin{equation}
  \label{eq:15}
f_i(z) \ = \ \sum_{l=0}^\infty \mathrm{E}\bigl[ N_i\,\mathbf{1}_{\{L = l\}}\bigr]\,z^l
\ = \ \mathrm{E}\bigl[N_i\,z^L\bigr]
\end{equation}
converge and can be represented as
\begin{equation}
  \label{eq:16}
f_i(z)\ = \
z^{\nu_i}\,g(z)\,\bigg( S_0\,r_{0,i} +
\sum_{j=1}^k \frac{\alpha_j\,\beta_j\,r_{j,i}}{1 +\beta_j - \beta_j\,\sum_{h=1}^n r_{j,h} z^{\nu_h}}\bigg).
\end{equation}
\end{subequations}
\end{theorem}
\textbf{Proof.} Note that $\mathrm{E}\bigl[L \,z^L\bigr] =
\sum_{l=0}^\infty l\,\mathrm{P}[L =l]\,z^l = z\, \frac{d g}{d
z}(z)$ is finite for $z \in (0,\rho_g)$ as a consequence of
general properties of power series. Since we assume $\nu_i > 0$, by
\begin{equation}
  \label{eq:32}
f_i(z) \ =\ \mathrm{E}\bigl[N_i\,z^L\bigr] \ \le \ \nu_i^{-1}\,\mathrm{E}\bigl[L \,z^L\bigr],
\end{equation}
also the $f_i(z)$ are finite for $z \in (0,\rho_g)$.
Fix $i \in\{1, \ldots, n\}$ and $z \in (0, \rho_g)$ and  observe that
\begin{equation}
\label{eq:8}
  \frac d {d\, h} \bigl( z^{L + h\,N_i}\bigr)\big|_{h = 0} \ = \
N_i\, z^L\, \log z.
\end{equation}
Recall from
section~\ref{sec:abstr-descr-probl} that we assume the $\nu_1,
\ldots, \nu_n$ to be positive integers.
Nevertheless, (\ref{eq:13}) holds not only for positive integers
$\nu_1, \ldots, \nu_n$ but more generally for any  $\nu_1, \ldots,
\nu_n
>0$. Thus, for small $h$ with $|h| < \nu_i$
\begin{equation}
  \label{eq:33}
\begin{split}
\mathrm{E}\bigl[ z^{L + h\,N_i}\bigr]& \ =\ \exp\left(S_0 \bigg(\,{\textstyle\sum\limits_{l=1, l\not=i}^n} r_{0,l}\,z^{\nu_l}
+r_{0,i}\,z^{\nu_i +h} -1\bigg)\right) \\
& \qquad \times \prod_{j=1}^k \left(1 +\beta_j - \beta_j\,\bigg(\,{\textstyle\sum\limits_{l=1, l\not=i}^n} r_{j,l}\, z^{\nu_l}
+r_{j,i}\,z^{\nu_i +h }\bigg)\right)^{-\alpha_j}.
\end{split}
\end{equation}
By (\ref{eq:33}), (\ref{eq:8}) will imply (\ref{eq:16}) as soon as it is clear that
the order of differentiation and expectation in
 $ \frac d {d\,h} \mathrm{E}\bigl[ z^{L + h\,N_i}\bigr]$
can be exchanged. But this follows from standard results on differentiation under the
integral \citep[e.g.][Theorem A.(9.1)]{D96}. \hfill $\Box$

Note that \citet[][translated to our notation]{MBT01} suggest the following approximation
for the risk contributions to VaR in the sense of (\ref{eq:7}):
\begin{subequations}
  \begin{equation}
    \label{eq:34}
\rho_i(L)\ \approx\ \nu_i\,\frac{\mathrm{E}[N_i\,\exp(s_\delta\,L)]}{\mathrm{E}[\,\exp(s_\delta\,L)]},
  \end{equation}
where $s_\delta>0$, the so-called \emph{saddle point}, is given as
the unique solution of the following equation
\begin{equation}
  \label{eq:35}
\frac{\mathrm{E}[L\,\exp(s_\delta\,L)]}{\mathrm{E}[\,\exp(s_\delta\,L)]}\ = \ \widehat{q}_\delta(L),
\end{equation}
with $\widehat{q}_\delta(L)$ standing either for $q_\delta(L)$ or any reasonable estimator of it.
\end{subequations}
Note that the right-hand side of (\ref{eq:34}) can be expressed as $\frac
1{s_\delta}\,\frac{d}{dh}\;\log \,{\mathrm{E}[\,\exp(z\,(L+h\,
N_i))]\bigg|_{(h=0,\,z=s_{\delta})}}$.

This approach is based on the so-called \emph{saddle-point method}
by means of which quantiles to tail probabilities of the
distribution of $L$ can be approximated \citep[see][]{MBT01,G01}.

Under Assumption \ref{as:2}, we obtain from Theorem~\ref{th:1} a
rather simple formula for the right-hand side of (\ref{eq:34}):
\begin{equation}
  \label{eq:36}
\rho_i(L)\ \approx\ \nu_i\,\exp(\nu_i\,s_\delta)\,\bigg( S_0\,r_{0,i} +
\sum_{j=1}^k \frac{\alpha_j\,\beta_j\,r_{j,i}}{1 +\beta_j - \beta_j\,\sum_{h=1}^n r_{j,h} \exp(\nu_h\,s_\delta)}\bigg).
\end{equation}
In order to derive the announced algorithm for the calculation of the VaR contributions in (\ref{eq:7}),
we need a further notation. Note that the probability measure $\mathrm{P}$ under consideration depends,
in particular,
on the parameters $\alpha_1, \ldots, \alpha_k$ from Assumption \ref{as:2}. We express this
dependence by writing
$$
\mathrm{P}\ =\ \mathrm{P}_\alpha \quad \mbox{with}\ \alpha = (\alpha_1,\ldots, \alpha_k),
$$
and, analogously, $\mathrm{E}_\alpha$ for the expectations.
\begin{corollary}
  \label{co:2}
Let $N_1, \ldots, N_n$ satisfy Assumption \ref{as:2}. Fix a confidence level $\delta \in (0,1)$ and calculate
the value-at-risk $q_\delta(L)$ of the portfolio loss $L$ according to (\ref{eq:4}) for the probability $\mathrm{P}_\alpha$.
Then, for $i=1, \ldots,n$, the VaR contributions in the sense of (\ref{eq:7}) can be calculated by means of
\begin{equation}
  \label{eq:18}
\mathrm{E}_\alpha[N_i\,|\, L = q_\delta(L)] \, = \, \frac{S_0\,r_{0,i}\,\mathrm{P}_{\alpha}[ L = q_\delta(L)-\nu_i] +
\sum_{j=1}^k \alpha_j\,\beta_j\,r_{j,i}\,
\mathrm{P}_{\alpha(j)}[ L = q_\delta(L) - \nu_i]}{\mathrm{P}_{\alpha}[ L = q_\delta(L)]},
\end{equation}
with $\alpha = (\alpha_1,\ldots, \alpha_k)$ and $\alpha(j) =(\alpha_1,\ldots,\alpha_j +1, \ldots, \alpha_k)$.
\end{corollary}
\textbf{Proof.} Fix any $i \in \{1, \ldots, n\}$. By
(\ref{eq:7a}), it suffices to show that the numerator of the
right-hand side of (\ref{eq:18}) equals
$\mathrm{E}_\alpha[N_i\,\mathbf{1}_{\{L = q_\delta(L)\}}]$. We
will prove the even stronger identity
\begin{equation}
  \label{eq:37}
\mathrm{E}_\alpha[N_i\,\mathbf{1}_{\{L = t\}}]\ =\
S_0\,r_{0,i}\,\mathrm{P}_{\alpha}[ L = t-\nu_i] + \sum_{j=1}^k
\alpha_j\,\beta_j\,r_{j,i}\, \mathrm{P}_{\alpha(j)}[ L = t -
\nu_i]\,,
\end{equation}
for any non-negative integer $t$. By Theorem \ref{th:1} we know
the generating function $f_i(z)$ of the sequence $t \mapsto
\mathrm{E}_\alpha[N_i\,\mathbf{1}_{\{L = t\}}]$. Write $g_\alpha$
instead of just $g$ in order to express the fact that $g$ like
$\mathrm{P}$ depends on the parameters $\alpha_1, \ldots,
\alpha_k$ from Assumption \ref{as:2}. With this notation,
(\ref{eq:16}) can be written as
\begin{equation}
  \label{eq:38}
  \begin{split}
f_i(z)& = z^{\nu_i}\,\bigg( S_0\,r_{0,i}\,g_\alpha(z) +
\sum_{j=1}^k \alpha_j\,\beta_j\,r_{j,i}\,g_{\alpha(j)}(z)\bigg)\\
& = \sum_{t=0}^\infty \bigg(S_0\,r_{0,i}\,\mathrm{P}_{\alpha}[ L =
t-\nu_i] + \sum_{j=1}^k \alpha_j\,\beta_j\,r_{j,i}\,
\mathrm{P}_{\alpha(j)}[ L = t - \nu_i] \bigg)\,z^t.
  \end{split}
\end{equation}
Note that for the second identity in (\ref{eq:38}) we have used
the fact that $L$ is non-negative. By the uniqueness of power
series, (\ref{eq:15}) and (\ref{eq:38}) imply now (\ref{eq:37}).
\hfill $\Box$

\begin{remark}\
  \label{rm:2}
\vspace{-2ex}
\begin{enumerate}
\nopagebreak[4]
\item Corollary \ref{co:2} states essentially that the VaR contributions
according to (\ref{eq:7}) can be determined by calculating the loss distribution $(k+1)$ times
with different parameters.
This is quite easy if the portfolio is parameterized as in Assumption~\ref{as:2}. But also in the case
described in Remark~\ref{rm:1}~(i) (with (\ref{eq:20}) as definition for $\tau_j$)
it is easy to run the algorithm in the appropriate parameterizations. Just note that in order to
calculate the $\mathrm{P}_{\alpha(j)}$ probabilities, it suffices to replace the weights
$\theta_{j,i}$, $i = 1, \ldots, n$, with $\theta'_{j,i} = \frac{\alpha_j + 1}{\alpha_j}\,\theta_{j,i}$.
\item By construction of $N_i$ as a conditionally Poisson distributed random variable, we
have $\mathrm{P}_\alpha[N_i > 1] > 0$.
Hence it is possible that the VaR contributions according to (\ref{eq:18}) and (\ref{eq:7}) become greater than the
exposures $\nu_i$.
\item A further consequence of the model construction according to Assumption \ref{as:2} is that
  \begin{equation}
    \label{eq:14}
\mathrm{E}_\alpha[N_i\,|\, L = q_\delta(L)] \ = \ 0
  \end{equation}
may happen. In fact, for fixed $i \in \{1, \ldots, n\}$ the following statement holds:\\
\emph{$\mathrm{E}_\alpha[N_i\,|\, L = q_\delta(L)]$ equals $0$ if and only if
$\sum\limits_{j=1}^n \nu_j\,u_j = q_\delta(L)$ with non-negative integers $u_1, \ldots, u_n$ implies
$u_i = 0$.
}\\
To see this, define $Y = L - \nu_i\,N_i$, and observe that
\begin{eqnarray}
\mathrm{P}[L = q_\delta(L) - \nu_i] = 0 & \iff&
\mathrm{P}[Y = q_\delta(L) - k\,\nu_i]=0,\ 1 \le k \le q_\delta(L)/\nu_i
\notag\\
& \iff& \mathrm{P}[L = q_\delta(L)] = \mathrm{P}[Y = q_\delta(L), N_i = 0]\notag\\
& \iff& \{ L = q_\delta(L)\} = \{ Y = q_\delta(L), N_i = 0\}.
 \label{eq:21}
\end{eqnarray}
Denote by $M$ the set of all $n$-tuples $(u_1, \ldots, u_n)$ of non-negative integers
such that\\
$\sum_{j=1}^n \nu_j\,u_j = q_\delta(L)$. Then the assertion follows from (\ref{eq:21}) and the fact
that
\begin{equation}
  \label{eq:22}
\{ L = q_\delta(L)\} \ =\ \bigcup_{(u_1, \ldots, u_n)\in M} \{N_1 = u_1, \ldots, N_n = u_n\}.
\end{equation}
In particular, (\ref{eq:14}) can occur in case $\nu_i > q_\delta(L)$. This is possible
whenever $p_i < 1 - \delta$ and $\nu_i$ is large compared to the other exposures.
\end{enumerate}
\end{remark}

Remark \ref{rm:2} (ii) is being caused by the fact that CreditRisk$^+$ makes use
of a Poisson approximation in order to compute the portfolio VaR. Hence one has to consider
the Poisson variables $N_i$ when computing risk contributions. As a consequence, there is little
hope that switching to another risk measure will solve the problem of contributions which are larger than
the corresponding exposures. The situation concerning Remark \ref{rm:2} (iii) is more favorable.

For instance, one can try to \emph{smooth} the conditional expectations in (\ref{eq:18}) by choosing
a fixed positive integer $t$ and computing $\mathrm{E}_\alpha[N_i\,|\, q_\delta(L)-t \le L \le  q_\delta(L)+t]$ instead of
$\mathrm{E}_\alpha[N_i\,|\, L = q_\delta(L)]$. We will not go into the details of this approach
since in the subsequent section we will see, that, in particular, the problem from Remark \ref{rm:2}~(iii)
can be avoided by switching
from VaR to Expected Shortfall, a risk measure which should be preferred for some
reasons \citep{AT01}.

\section{Contributions to Expected Shortfall}
\label{sec:contr-expect-shortf}

The \emph{Expected Shortfall (ES)} at level $\delta\in (0,1)$ of the portfolio loss $L$ can be defined as
\begin{equation}
  \label{eq:23}
\mathrm{ES}_\delta(L) \ = \ (1-\delta)^{-1} \int_\delta^1 q_u(L)\,d u.
\end{equation}
It may be characterized as the smallest \emph{coherent} risk measure dominating VaR and only
depending on $L$ through its distribution \citep[][Th.~6.10]{Delb98}.
Of course, (\ref{eq:23}) appears not very handy for calculations or defining ES contributions.
Nevertheless, if $L$ had a continuous distribution, the representation
\begin{equation}
  \label{eq:24}
\mathrm{ES}_\delta(L) \ = \ \mathrm{E}[L\,|\, L \ge q_\delta(L)]
\end{equation}
would be equivalent to (\ref{eq:23}) \citep[][Cor.~5.3]{AT01}. Indeed, if we assume to deal
with a large portfolio we can hope that the difference between (\ref{eq:23}) and (\ref{eq:24}) will not
be too large. With (\ref{eq:24}) as definition, the decomposition
\begin{equation}
  \label{eq:25}
\mathrm{E}[L\,|\, L \ge q_\delta(L)] \ = \ \sum_{i=1}^n \nu_i\,\mathrm{E}[N_i\,|\, L \ge q_\delta(L)]
\end{equation}
suggests the choice of $\nu_i\,\mathrm{E}[N_i\,|\, L \ge q_\delta(L)]$ as ES contribution of asset $i$.
Further evidence for this choice stems from the fact under certain smoothness assumptions (\ref{eq:25})
is just the decomposition corresponding to (\ref{eq:5}) with $\rho = \mathrm{ES}$ \citep[][Lemma~5.6]{T99}.

From Corollary \ref{co:2} we obtain the following result on the computation of the
ES contributions.
\begin{corollary}
  \label{co:3}
Let $N_1, \ldots, N_n$ satisfy Assumption \ref{as:2}. Fix a confidence level $\delta \in (0,1)$ and calculate
the value-at-risk $q_\delta(L)$ of the portfolio loss $L$ according to (\ref{eq:4}) for the probability $\mathrm{P}_\alpha$.
Then, for $i=1, \ldots,n$, the ES contributions in the sense of (\ref{eq:25}) can be calculated by means of
\begin{equation}
  \label{eq:26}
\mathrm{E}_\alpha[N_i\,|\, L \ge q_\delta(L)] \, = \,\frac{S_0\,r_{0,i}\,\mathrm{P}_{\alpha}[ L \ge q_\delta(L)-\nu_i] +
\sum_{j=1}^k \alpha_j\,\beta_j\,r_{j,i}\,
\mathrm{P}_{\alpha(j)}[ L \ge q_\delta(L) - \nu_i]}{\mathrm{P}_{\alpha}[ L \ge q_\delta(L)]},
\end{equation}
with $\alpha = (\alpha_1,\ldots, \alpha_k)$ and $\alpha(j) =(\alpha_1,\ldots,\alpha_j +1, \ldots, \alpha_k)$.
\end{corollary}
\textbf{Proof.} Observe that
\begin{equation}
  \label{eq:27}
\mathrm{E}_\alpha[N_i\,|\, L \ge q_\delta(L)] \ = \
\frac{\sum_{t=q_\delta(L)}^\infty \mathrm{E}_\alpha[N_i\,\mathbf{1}_{\{L=t\}}]}{\mathrm{P}_\alpha[ L \ge q_\delta(L)]}
\end{equation}
and note that in the proof of Corollary \ref{co:2} we have in fact shown that
\begin{equation}
  \label{eq:30}
\mathrm{E}_\alpha[N_i\,\mathbf{1}_{\{L=t\}}]\,=\,
S_0\,r_{0,i}\,\mathrm{P}_{\alpha}[ L = t-\nu_i] +
\sum_{j=1}^k \alpha_j\,\beta_j\,r_{j,i}\,
\mathrm{P}_{\alpha(j)}[ L = t - \nu_i]
\end{equation}
for any non-negative integer $t$.
\hfill $\Box$

Note that none of the probabilities in the numerator of the right-hand side of (\ref{eq:26}) can become $0$,
since by definition of $q_\delta(L)$ the probability $\mathrm{P}_\alpha[L \ge q_\delta(L)-\nu_i]$ satisfies
\begin{subequations}
  \begin{equation}
    \label{eq:28}
\mathrm{P}_\alpha[L \ge q_\delta(L)-\nu_i]\ \ge\ \mathrm{P}_\alpha[L \ge q_\delta(L)]\  \ge \ 1 -\delta
  \end{equation}
and for any integer $t$ we have
\begin{equation}
  \label{eq:29}
\mathrm{P}_\alpha[L = t] = 0 \quad \iff \quad \mathrm{P}_{\alpha(j)}[L = t] = 0,\ j = 1, \ldots, k.
\end{equation}
\end{subequations}
Nonetheless, the observation in Remark \ref{rm:2} (ii) holds for the ES contributions as well.

\section{Conclusion}

Attributing the total risk of a credit portfolio to its components in an exhaustive
and risk respecting way is an important task in portfolio management.
For the case that risk is  measured as value-at-risk (VaR) and determined with
the CreditRisk$^+$ model \citep{CRplus}, we have shown that this attribution
can be performed by calculating the loss distribution $(k+1)$ times ($k$ denoting the number
of sectors in the model) with slightly different parameters. The same statement holds for the attribution when risk is
measured as Expected Shortfall (ES).


\end{document}